\documentclass[12pt]{article}
\usepackage{amsfonts}
\usepackage{amsmath, amssymb}
\usepackage{hyperref}
\usepackage[dvipdf]{color}
\usepackage{graphicx, subfigure}

\usepackage{color}

\renewcommand{\title}[1]{\vbox{\center\LARGE{#1}}\vspace{5mm}}
\renewcommand{\author}[1]{\vbox{\center\large#1}\vspace{5mm}}
\newcommand{\address}[1]{\vbox{\center\em#1}}

\numberwithin{equation}{section}

\newcommand{\beq}{\begin{equation}}
\newcommand{\eeq}{\end{equation}}
\newcommand{\bea}{\begin{eqnarray}}
\newcommand{\eea}{\end{eqnarray}}

\begin{document}

\bibliographystyle{utphys}

\begin{titlepage}
\begin{center}

\vspace{5mm}
\hfill {\tt UT-Komaba/13-3, CALT-68-2923}\\
\vspace{16mm}

\title{
Scheme dependence of instanton counting\\
 in ALE spaces
}
\vspace{6mm}

Yuto Ito${}^{\spadesuit}$\footnote{\href{mailto:ito@hep1.c.u-tokyo.ac.jp}
{\tt ito@hep1.c.u-tokyo.ac.jp}},
Kazunobu Maruyoshi${}^{\heartsuit}$\footnote{\href{mailto:maruyosh@caltech.edu}
{\tt maruyosh@caltech.edu}},
and
Takuya Okuda${}^{\spadesuit}$\footnote{\href{mailto:takuya@hep1.c.u-tokyo.ac.jp}
{\tt takuya@hep1.c.u-tokyo.ac.jp}},
\vskip 5mm
\address{
${}^\spadesuit$%
University of Tokyo, Komaba\\
 Meguro-ku, Tokyo 153-8902, Japan
}
\address{ 
${}^\heartsuit$California Institute of Technology\\
452-48, Pasadena, California 91125, USA
}

\end{center}

\vspace{8mm}
\abstract{
\noindent
There have been two distinct schemes studied in the literature for instanton counting in $A_{p-1}$ asymptotically locally Euclidean (ALE) spaces.
We point out that the two schemes---namely the counting of orbifolded instantons and 
instanton counting in the resolved space---lead in general to different results for partition functions.
We illustrate this observation in the case of ${\cal N}=2$ $U(N)$ gauge theory with $2N$ flavors on the $A_{p-1}$ ALE space.
We propose simple relations between the instanton partition functions given by the two schemes and test them by explicit calculations.
}
\vfill

\end{titlepage}

\setcounter{footnote}{0}


\section{Introduction}

In this paper we make an observation on the two schemes used in the literature for instanton counting in $A_{p-1}$ asymptotically locally Euclidean (ALE) spaces.
The first scheme uses the enumeration of torus fixed points in the moduli space of instantons on $\mathbb C^2$ \cite{Nekrasov:2002qd}, and keeps only contributions that are $\mathbb Z_p$-invariant in the sense we will explain.
The second is based on such enumeration of fixed points in the moduli space of instantons on the resolved $A_{p-1}$-ALE space.
Since this space is the minimal resolution of the orbifold $\mathbb C^2/\mathbb Z_p$, one naturally expects that the results of the two counting schemes are simply related.
In fact, in the examples studied in the literature, the two schemes produce identical results.

We point out that in general the two schemes lead to different results.
Our experience shows that the difference appears 
when there are a sufficient number of fundamental/anti-fundamental/bifundamental hypermultiplets 
{\it and} when the sectors with non-zero values of the first Chern class are considered.
As far as we know, the appearance of the difference has not been noticed in the literature.

We illustrate our observation by calculating the instanton partition functions of the ${\mathcal N}=2$ $U(N)$ gauge theory with $N_\text{F}=2N$ flavors, {\it i.e.}, $N$ fundamental and $N$ anti-fundamental hypermultiplets.
In \S \ref{sec:orbifold}, we apply the counting of orbifolded instantons \cite{MR1075769,Fucito:2004ry} 
and obtain the instanton partition function.
We then consider instanton counting in the resolved spaces in \S \ref{sec:resolved}.
In \S \ref{sec:A1-resolved} we first focus on the resolved $A_1$-ALE space since  the instanton counting scheme for this space has been rigorously established \cite{Nekrasov:2003vi,Sasaki:2006vq,2009CMaPh.tmp..240G,2011CMaPh.304..395B,Bonelli:2011jx,Bonelli:2011kv}.%
\footnote{%
See \cite{Cirafici:2012qc} for a review and references on instanton counting in toric spaces.
}
In \S \ref{sec:A-general-resolved} we analyze the resolved $A_{p-1}$ spaces with general $p\geq 2$ by applying the physically motivated method developed in \cite{Bonelli:201208}. 
In \S \ref{sec:proposals} we propose simple relations between the instanton partition functions that result from the two schemes.

Our study of instanton counting in ALE spaces was motivated by a version of the AGT correspondence \cite{AGT}.
It was found in \cite{Belavin:2011pp} that the norm of the Whittaker vector in the super Virasoro algebra is identical to the instanton partition function of the pure $SU(2)$ theory on $\mathbb C^2/\mathbb Z_2$.
The super Virasoro algebra with a generic central charge is realized by super Liouville theory.
The correspondence between super Liouville theory and $\mathcal N=2$ theories on the $A_1$-ALE space has been extended in many directions; in particular the gauge theory calculations were performed in various settings in \cite{
Bonelli:2011jx,Belavin:2011tb,Bonelli:2011kv,Wyllard:2011mn,Ito:2011mw,Alfimov:2011ju,Belavin:2011sw,Krefl:2011aa,Belavin:2012uf,Bonelli:201208,Belavin:2012aa,Belavin:2012eg}.
It was noticed in \cite{Bonelli:2011jx,Bonelli:2011kv} that the instanton partition function computed on the resolved space has a structure that naturally matches two copies of Liouville theory.
This was studied in detail in \cite{Belavin:2011sw}.
We discuss implications of our observation for the correspondence with 2d theories in \S \ref{sec:discussion}.
We also note there that the results in this paper have useful applications for the computation of 't Hooft line operators.

In the appendix we collect the details of the calculations that support our proposals.

\section{Counting orbifolded instantons}
\label{sec:orbifold}

In this section, we apply the counting scheme based on orbifolded instantons to the ${\mathcal N}=2$ $U(N)$ gauge theory with $N$ fundamental and $N$ anti-fundamental hypermultiplets.
This scheme is based on \cite{MR1075769} and was developed in \cite{Fucito:2004ry}.
(See also \cite{Fucito:2006}.)
 We denote the scalar vevs by $\vec a = (a_1, \cdots, a_N)$, anti-fundamental masses by $\mu_1,\cdots, \mu_N$, and fundamental masses by $\mu_{N+1},\cdots, \mu_{2N}$.
We set $\boldsymbol \mu=(\mu_{1},\cdots, \mu_{2N})$.

The asymptotic boundary of the $A_{p-1}$ space is the lens space $\mathbb S^3/\mathbb Z_p$, which has non-contractible torsion 1-cycles.
Thus the gauge field can have a nontrivial holonomy 
\begin{equation}
U=\text{diag}( e^{2\pi i  I_1/p} , e^{2\pi i  I_2/p} , \cdots, e^{2\pi i I_N/p}  )  
\end{equation}
along the generator of $\pi_1(\mathbb S^3/\mathbb Z_p)=\mathbb Z_p$, where $I_\alpha\in \mathbb{Z}$ and $I_\alpha\sim I_\alpha+p$.
We use the notation $\vec I=(I_1,\ldots,I_N)$.

The instanton partition function on $\mathbb C^2/\mathbb Z_p$ can be obtained by summing the $\mathbb Z_p$-invariant contributions of the torus fixed points in the moduli space of instantons on $\mathbb C^2$.
The $\mathbb Z_p$-action depends on $\vec I$.
We can label the fixed points by the same $N$-tuples of Young diagrams $\vec{Y}=(Y_1, \cdots, Y_N)$ used in the instanton counting calculations on $\mathbb{C}^2$, which we review in \S\ref{app:C2}.

Although we work with the singular orbifold, it is useful to keep track of the first Chern class $c_1(E)$ of the gauge bundle $E$ that becomes well-defined once
 the space gets resolved.
We decompose it as 
\begin{equation}
c_1(E)=\sum_{r={ 1}}^{p-1} c_{(r)}  c_1(T_r)\,,
\end{equation}
where $T_r$ is the flat line bundle with holonomy $e^{2\pi i r/p}$. 
Let us introduce the notation 
\begin{equation}
\boldsymbol{c} \equiv (c_{(1)}, \cdots, c_{(p-1)})\in\mathbb Z^{p-1}\,.
\end{equation}
The precise range of $\boldsymbol c$ depends on $\vec I$.

The combined data ($\vec I$, $\vec Y$) uniquely determines $\boldsymbol c$.
The map $(\vec I,\vec Y)\mapsto \boldsymbol c$ is given as follows.
Given the holonomy data $\vec I$ we assign the additive $\mathbb{Z}_p$ charge $I_\alpha + (i-1)-(j-1)$ mod $p$ to the box in the $i$-th column and the $j$-the row of the Young diagram $Y_\alpha$.
Denote by $N_r$ the number of elements with value $e^{2\pi i  r/p}$ in $U$.
Denote also by $K_r$ the number of boxes with charge $r$ mod $p$ in all the diagrams in $\vec Y$.
Each $\vec Y$ contributes to the partition function with the first Chern class 
given by
\begin{equation}\label{fucito}
c_{(r)}=  N_r-2K_r+K_{r+1}+K_{r-1}\,,\qquad
r=1,\ldots,p-1
\,.
\end{equation}
For a given pair $(\vec{I}, \boldsymbol{c})$, there are infinitely many $N$-tuples $\vec{Y}$ of Young diagrams that satisfy the relation (\ref{fucito}).
We denote the set of such $\vec{Y}$ by $\mathcal Y(\vec I,\boldsymbol c)$.

Recall that the instanton partition function for $\mathbb C^2$ takes the form
 \begin{align}
Z^{\mathbb{C}^2}_{N_{\rm F}=2N, \,  {\rm inst}}(\vec{a}; {\boldsymbol \mu} 
;q;\boldsymbol \epsilon)=\displaystyle\sum_{\vec{Y}} q^{|\vec{Y}|} 
Z^{\mathbb{C}^2}_{N_{\rm F}=2N, \, \vec{Y}}(
\vec{a}; {\boldsymbol \mu} ; \boldsymbol \epsilon  ) \, ,
\end{align}
where $Z^{\mathbb{C}^2}_{N_{\rm F}=2N, \, \vec{Y}}$ in each term is the product of the weights of the equivariant $\mathbb T$-action.
The associated equivariant parameters are $\vec a,\boldsymbol \mu$, and $\boldsymbol\epsilon=(\epsilon_1,\epsilon_2)$.
The group $\mathbb T=(\mathbb{C}^*)^N \times (\mathbb{C}^*)^{2N} \times (\mathbb{C}^*)^2$ is the complexified maximal torus of $G\times G_\text{F}\times SO(4)$, where $G=U(N)$ is the gauge group, $G_\text{F}=U(N)\times U(N)$ is the flavor group, and $SO(4)$ is the Lorentz group.
The explicit expression for $Z^{\mathbb{C}^2}_{N_{\rm F}=2N, \, \vec{Y}}$ can be found in \S \ref{app:C2}.
In the case of $\mathbb{C}^2/\mathbb{Z}_p$ the product must be restricted to $\mathbb Z_p$-invariant weights, {\it i.e.}, the weights that have a vanishing $\mathbb{Z}_p$ charge. 
The Coulomb vev $a_{\alpha}$ has charge  $I_{\alpha}$, and $(\epsilon_1,\epsilon_2)$ have charges $(1,-1)$.
The instanton partition function with holonomy $\vec{I}$ and the first Chern class $\boldsymbol{c}$ on $\mathbb{C}^2/\mathbb{Z}_p$ is given as
\begin{align}
\label{orb}
Z^{\mathbb{C}^2/\mathbb{Z}_p}_{N_{\rm F}=2N,\,{\rm inst},\, \boldsymbol{c} \, }(\vec{a}, \vec{I} ;\boldsymbol \mu;q;\boldsymbol\epsilon)=\sum_{\vec Y \in 
\mathcal Y(\vec I,\boldsymbol c)
 } q^{
K_0 +\frac{1}{2} \sum_{r=0}^{p-1} C_r N_r
} 
Z^{\mathbb{C}^2/\mathbb{Z}_p}_{N_{\rm F}=2N,\, \vec Y}(\vec{a},\vec{I}; \boldsymbol \mu; \boldsymbol\epsilon)\, ,
\end{align}
where $C_r=\frac{1}{p}(p-r)r$.
The factor $Z^{\mathbb{C}^2/\mathbb{Z}_p}_{N_{\rm F}=2N,\, \vec Y}$ is defined in the same way as $Z^{\mathbb{C}^2}_{N_{\rm F}=2N, \, \vec{Y}}$, except that the products in (\ref{z-vec}-\ref{z-anti-fund}) are restricted to the invariant weights.
We present the explicit calculation up to several orders of $q$ in \S\ref{app:orbifold}.

\section{Instanton counting in the resolved spaces}
\label{sec:resolved}

In this section we review and apply the second scheme for instanton counting in $A_{p-1}$-ALE spaces.
Physically, the idea can be summarized as follows.
Upon performing the orbifolding by $\mathbb Z_p$ and the minimal resolution, the maximal torus $U(1)^2$ of the Lorentz group $SO(4)$ descends to an isometry of the resolved $A_{p-1}$ space.
The resolution also produces homologically non-trivial submanifolds each isomorphic to $\mathbb P^1$.
The $\mathbb P^1$ submanifolds intersect with each other at their north and south poles.
The torus action has $p$ fixed points precisely at the poles.
The instanton partition function on the resolved $A_{p-1}$ space is obtained by gluing the instanton contributions from the $p$ fixed points, taking into account also the bulk contributions to the fluctuation determinant.

The $p=2$ case is mathematically more rigorous; 
the Poincar\'e polynomials were computed in \cite{2011CMaPh.304..395B}, while \cite{Bonelli:2011jx,Bonelli:2011kv} adapted the method for the calculation of the instanton partition functions with the vanishing first Chern classes.
For general $p$, we use the method proposed in \cite{Bonelli:201208}.
When specialized to $p=2$, the latter method reproduces the results from the first one.

\subsection{$A_1$-ALE space}
\label{sec:A1-resolved}

Here we consider the $p=2$ case and denote the first Chern class $c_{(1)}$ by $c$. 
The supersymmetric saddle point configurations in the path integral are abelian and can be diagonalized.
Such configurations can be partially classified by the first Chern class of each $U(1)$ factor in the unbroken gauge group.
We parametrize the first Chern class of the $\alpha$-th $U(1)$ subgroup by $k_\alpha \in \frac{1}{2} \mathbb{Z}$.
The normalization is such that $2\sum_{\alpha=1}^{N} k_{\alpha}=-c$.

The $A_{1}$-ALE space has a $\mathbb{P}^{1}$ by which the orbifold singularity are blown up.
As mentioned above, the instanton partition function on this space is obtained 
by intertwining the contributions from two fixed points at the north and south poles of the $\mathbb{P}^{1}$,
multiplied by the so-called $\ell$-factor which will be introduced shortly.
Since each fixed point has a neighborhood locally isomorphic to $\mathbb{C}^{2}$,  its contribution is simply the instanton partition function on $\mathbb{C}^{2}$ \cite{Nekrasov:2002qd}.
The weights $\epsilon^{(0)}_{1,2}$ and $\epsilon^{(1)}_{1,2}$ of the torus action $U(1)^{2}$ on the local invariant coordinates at the north and south poles are given by
  \bea
  \epsilon^{(0)}_1=2\epsilon_1,~~  \epsilon^{(0)}_2=-\epsilon_1+\epsilon_2, ~~~
  \epsilon^{(1)}_1=\epsilon_1-\epsilon_2, ~~~ \epsilon^{(1)}_2=2\epsilon_2.
  \eea 
Furthermore, at these poles the scalar vev $a_{\alpha}$ gets shifted to 
(see \cite{Bonelli:201208} for an explanation)
  \bea
  a_{\alpha}^{(0)} = a_{\alpha} -2\epsilon_1 k_{\alpha} \, ,& 
  a_{\alpha}^{(1)} = a_{\alpha} -2\epsilon_2 k_{\alpha} \, .
  \label{ae}
  \eea

Let us introduce the $\ell$-factors as follows.
First we define
\begin{equation}
    \begin{aligned}
  &  \ell_{\alpha \beta}(x, k_{\alpha};\tilde{x}, \tilde{k}_{\beta};m; \boldsymbol \epsilon)
\\
&     =     \left\{ \begin{array}{ll}
           \displaystyle{\prod_{{\scriptstyle i,j\ge 0, \ i+j \le 2(k_{\alpha \beta} - 1)} 
           \atop {\scriptstyle i+j-2k_{\alpha \beta} \equiv 0 \ {\rm mod} ~2}}
           (x - \tilde{x} - i \epsilon_1 - j \epsilon_2 - m) } , & k_{\alpha \beta} > 0\,,   \\
           \displaystyle{ \prod_{{\scriptstyle i,j\ge 0, \ i+j \le - 2(k_{\alpha \beta}+1)} 
           \atop {\scriptstyle i+j-2k_{\alpha \beta} \equiv 0 \ {\rm mod} ~2}}
           (x - \tilde{x} + (i+1) \epsilon_1 + (j+1) \epsilon_2 - m)}, & k_{\alpha \beta} < 0 \,,  \\
           1 , &  k_{\alpha \beta} = 0\,,
           \end{array} \right.
             \end{aligned}
\nonumber
\end{equation}
where $k_{\alpha}, \tilde{k}_{\beta}\in(1/2) \mathbb{Z}$, $1 \leq \alpha,\beta \leq N$, and $k_{\alpha\beta}:=k_{\alpha}-\tilde{k}_{\beta}$.
  Similarly, we define
    \begin{align}
    \ell_{\alpha}(x, k_{\alpha};\boldsymbol \epsilon)
     =     \left\{ \begin{array}{ll}
           \displaystyle{\prod_{{\scriptstyle i,j\ge 0,\, \ i+j \le 2 (k_\alpha - 1)} 
           \atop {\scriptstyle i+j+ 2 k_\alpha \equiv 0 \ {\rm mod}~ 2}}
           (- x + (i+1) \epsilon_1 + (j+1) \epsilon_2 )} , &\quad k_{\alpha} > 0\,,  \\
           \displaystyle{\prod_{{\scriptstyle i,j\ge 0, \ i+j \le - 2 (k_{\alpha}+1)}
           \atop {\scriptstyle i+j + 2 k_\alpha \equiv 0 \ {\rm mod}~ 2}}
           (- x - i \epsilon_1 - j \epsilon_2)}, &\quad k_{\alpha} < 0\,,   \\
           1 , &\quad  k_{\alpha} = 0\,.
           \end{array} \right.\nonumber
    \end{align}
With these definitions, the $\ell$-factors for the bifundamental and (anti-)fundamental hypermultiplets
  are given by
    \begin{align}
    \ell^{A_1{\rm \mathchar`-ALE}}_{{\rm bifund}}(\vec{a}, \vec{k}; \vec{\tilde{a}}, \vec{\tilde{k}};m;\boldsymbol \epsilon)
&    =    \prod_{\alpha = 1}^{{ N}} \prod_{\beta = 1}^{{ N}}
           \ell_{\alpha \beta} (a_{\alpha}, k_{\alpha}; \tilde{a}_{\beta}, \tilde{k}_{\beta}; m;\boldsymbol \epsilon)\,,
         \nonumber   \\
    \ell^{A_1{\rm \mathchar`-ALE}}_{{\rm fund}}(\vec{a},\vec{k};m;\boldsymbol \epsilon)
&    =    \prod_{\alpha} \ell_{\alpha} (a_{\alpha} + \epsilon_+ - m, k_\alpha;\boldsymbol \epsilon)\, ,\nonumber\\
        \ell^{ A_1{\rm \mathchar`-ALE}}_{{\rm anti\mathchar`-fund}} (\vec{a},\vec{k};m;\boldsymbol \epsilon)
&     =     \prod_{\alpha} \ell_{\alpha} (a_{\alpha} + m, k_\alpha;\boldsymbol \epsilon)\,,
           \nonumber
    \end{align}
  while the $\ell$-factors for the adjoint hypermultiplet and the vector multiplet are given by
    \begin{equation}
    \begin{aligned}
   \ell^{ A_1{\rm \mathchar`-ALE}}_{{\rm adj}}(\vec{a},\vec{k};m;\boldsymbol \epsilon)
&     =     \ell^{A_1{\rm \mathchar`-ALE}}_{{\rm bifund}}(\vec{a}, \vec{k}; \vec{a}, \vec{k};m;\boldsymbol \epsilon)\,, 
\\
    \ell^{A_1{\rm \mathchar`-ALE}}_{{\rm vector}} (\vec{a},\vec{k};\boldsymbol \epsilon)
&     =     \ell^{ A_1{\rm \mathchar`-ALE}}_{{\rm adj}} (\vec{a},\vec{k};0;\boldsymbol \epsilon)^{-1}\,. 
    \end{aligned}
    \end{equation}
We then define the total $\ell$-factor of $U(N)$ gauge theory with $N_{f} =2N$ by taking a product over all multiplets:
\begin{align}\label{eq:ell-total-A1}
\ell^{A_1{\rm \mathchar`-ALE}}_{N_{\text F}=2N}( \vec{a},\vec{k};{\boldsymbol \mu};{\boldsymbol \epsilon})=\ell^{ A_1{\rm \mathchar`-ALE}}_{\rm vector}(\vec{a},\vec{k};{\boldsymbol \epsilon})
\prod_{i=1}^N \ell^{ A_1{\rm \mathchar`-ALE}}_{\rm anti \mathchar`- fund}(\vec{a},\vec{k};\mu_i;{\boldsymbol \epsilon})
\prod_{j=N+1}^{2N} \ell^{A_1{\rm \mathchar`-ALE}}_{\rm fund}(\vec{a},\vec{k};\mu_j;{\boldsymbol \epsilon}) \, .
\end{align}

The instanton partition functions with holonomy $\vec{I}$ and the first Chern class $c$ on the resolved $A_1$-ALE space is given as
\begin{align}
&Z^{ A_1{\rm \mathchar`-resolved}}_{N_{\text F}=2N,\, {\rm inst } ,\, c}(\vec{a}, \vec{I};{\boldsymbol \mu};q;{\boldsymbol \epsilon} )\nonumber\\
\label{A1ALEinst}
&\hspace{3em}=
\sum_{\vec k\in    \mathcal K(\vec I,\boldsymbol c)}
q^{ \sum^N_{\alpha=1} k^2_{\alpha} } \,
\ell^{ A_1{\rm \mathchar`-ALE}}_{N_{\text F}=2N}( \vec{a},\vec{k};{\boldsymbol \mu};{\boldsymbol \epsilon})
 \prod_{r=0,1} Z^{\mathbb{C}^2}_{N_{\text F}=2N,\, {\rm inst}} (\vec{a}^{(r)}; {\boldsymbol \mu};q;\epsilon^{(r)}_1, \epsilon^{(r)}_2) \, ,
\end{align}
where
 \begin{equation}
   \mathcal K(\vec I,\boldsymbol c)=
\left\{
\left. \vec k \in \left(\frac12\mathbb{Z}\right)^N 
\right
|
2\sum_{\alpha} k_{\alpha}=- c\,, \ e^{2\pi i k_{\alpha}}= e^{2\pi i I_{\alpha}/2}  
\text{ for } 1\leq\alpha\leq N
\right\}\,,
 \end{equation}
and the instanton partition function on $\mathbb{C}^2$, denoted by $Z^{\mathbb{C}^2}_{N_{\text F}=2N,\, {\rm inst}}$, is provided in (\ref{eq:ZY-C2}). 
We give an explicit calculation up to a few orders of $q$ in \S\ref{app:stacky}.

\subsection{$A_{p-1}$-ALE space}
\label{sec:A-general-resolved}

We now turn to the general $A_{p-1}$ case.
The saddle point configurations are again abelian, and the gauge bundle decomposes into $N$ line bundles: $E=\oplus_{\alpha=1}^N L_\alpha$.
In the resolved space, it is natural to use by Poincar\'e duality the exceptional divisors $\Sigma_r$ ($r=1,\ldots,p-1$) as a basis of the second (co)homology.
We expand $c_1(L_\alpha)=-\sum_{r=1}^{p-1} k^{(r)}_\alpha \Sigma_r$ with the coefficients $k^{(r)}_\alpha$ taking values in $(1/p)\mathbb Z$.
The basis $\{\Sigma_r\}$ is dual to the basis $\{c_1(T_r)\}$ we used in \S\ref{sec:orbifold} \cite{MR1075769}.
The coefficients are related as
$c_{(r)}=\sum_{s,\alpha} C_{rs} k^{(s)}_{\alpha}$, where $C_{rs}$ is the Cartan matrix:
$C_{rr}=2$ ($r=1,\ldots,p-1$), $C_{r,r+1}=C_{r+1,r}=-1$ ($r=1,\ldots,p-2$), and the other elements vanish.
We use the notation $\boldsymbol{\vec k}=(\vec k^{(1)},\ldots,\vec k^{(p-1)})$ and 
$\boldsymbol{c}=(c_{(1)},\ldots,c_{(p-1)})$.

Let us review the method proposed in \cite{Bonelli:201208}.
The total partition function on the resolved ALE space splits into the classical, one-loop, and instanton parts.
The one-loop part is the fluctuation determinant in the topologically trivial background, and should be universal in all topological sectors once the asymptotic boundary condition is fixed by $\vec I$.
Assuming that at least some sectors have the same partition functions as computed by the orbifold method in \S\ref{sec:orbifold}, we can compute the one-loop determinant by restricting to the $\mathbb Z_p$-invariant factors 
of the $\mathbb C^2$ one-loop factor by using the orbifold method explained in \S \ref{sec:orbifold}.

We expect that the 
total partition function for fixed $\boldsymbol{\vec k}$ precisely factorizes into the contributions from the $p$ fixed points in the ALE space, each of which can be written as the total partition function on $\mathbb{C}^{2}$ \cite{Nekrasov:2003vi}.
Such factorization is expected  because the total fluctuation determinant for each saddle point configuration should be calculable 
by the localization formula for the equivariant index of appropriate differential operators in the non-compact case.
Examples include the blow-up of $\mathbb{P}^{2}$  \cite{2003math.....11058N}, and the $A_{1}$-ALE space \cite{Bonelli:2011kv} above.
Explicitly, for the resolved space $A_{p-1}$ we expect the relation
  \bea
  Z_{{\rm total},\boldsymbol{c}}^{A_{p-1}{\rm \mathchar`-resolved}}(\vec a, \vec{I}; {\boldsymbol{\mu}}; q; \boldsymbol{\epsilon})
   =     \sum_{{\vec {\boldsymbol{k}}}\in \mathcal K(\vec I,\boldsymbol{c})} 
         \prod_{r=0}^{p-1} Z^{\mathbb{C}^{2}}_{{\rm total},\boldsymbol{c}}
(\vec{a}^{(r)}, \boldsymbol{\mu};q; \epsilon^{(r)}_{1}, \epsilon^{(r)}_{2})\,.
         \label{total}
  \eea
Here the index $r$ labels the fixed points and 
the equivariant parameters $\epsilon^{(r)}_{1,2}$ of the torus action and shifts of the vevs are
  \begin{align}
  \begin{array}{ll}
  \epsilon^{(r)}_1= (p-r) \epsilon_1 -r \epsilon_2 ,  \qquad  \epsilon^{(r)}_2 =  (-p+r +1) 
  \epsilon_1+ (r+1) \epsilon_2 \, ,\\
  \qquad\qquad\qquad
  a_{\alpha}^{(r)} = a_{\alpha} + k_{\alpha}^{(r+1)}  \epsilon^{(r)}_1  + k_{\alpha}^{(r)}  \epsilon^{(r)}_2  \, .
  \end{array} \label{Ap-1epsilon}
  \end{align}
We also defined set $\mathcal K(\vec I,\boldsymbol{c})$ as
\begin{equation}\label{K-def-p-general}
   \begin{aligned}
   \mathcal K(\vec I,c_{(s)})
&\equiv
\Bigg\{ 
\vec k^{(r)} \in \left( \frac{1}{p} \mathbb{Z} \right)^{N(p-1)}\, \Bigg|  \
\sum_{t=1}^{p-1} C_{rt} k^{(t)}_{\alpha} \in \mathbb{Z}\,,
\\
&
\qquad\qquad\qquad\qquad
c_{(s)}=\sum_{r,\alpha} C_{sr} k^{(r)}_{\alpha}, \ e^{-2\pi i k^{(1)}_{\alpha}}= e^{2\pi i I_{\alpha}/p}
\Bigg\}\,.
   \end{aligned}
 \end{equation}
The relation (\ref{total}) specialized to $p=2$ implies the equality (\ref{A1ALEinst}), with the arguments shifted as in \eqref{Ap-1epsilon}, and with the $\ell$-factors given as the ratio of the $\mathbb Z_2$-orbifolded one-loop factor and the product of two one-loop factors on $\mathbb C^2$.
The resulting $\ell$-factors are precisely those given in (\ref{eq:ell-total-A1}).
The authors of \cite{Bonelli:201208} proposed that this can be generalized to arbitrary $p\geq 2$;
one can obtain the $\ell$-factors by computing the ratio of the $\mathbb Z_p$-orbifolded one-loop factor and the product of $p$ one-loop factors on $\mathbb C^2$.

In order to write down the explicit one-loop contributions, let us introduce the functions%
\footnote{%
Our convention for $\gamma_{\epsilon_1,\epsilon_2}(x)$ agrees with \cite{Nekrasov:2003rj,2003math.....11058N,AGT} and differs from the one in \cite{Bonelli:201208}.
}
\begin{align}
& \gamma_{\epsilon_1,\epsilon_2}(x)
:=
         \frac{d}{ds}\Big|_{s=0} \frac{
         1}{\Gamma(s)}
         \int_0^\infty \frac{dt}{t}
           t^s \frac{e^{-tx}}{(e^{\epsilon_1t}-1)(e^{\epsilon_2t}-1)}\, , \nonumber\\
 & \gamma^{(p)}_{\epsilon_1,\epsilon_2}(x,n):=
\sum_{r=0}^{p-1}
\gamma_{p\epsilon_1, p\epsilon_2}( x - r\epsilon_1- [-pn+r]_p\, \epsilon_2) \, ,  \nonumber
\end{align}
where $ n\in (1/p) \mathbb{Z}$. 
The symbol $[x]_p$ for $x\in\mathbb Z$ denotes the integer that satisfies $0\leq [x]_p \leq p-1$ and $[x]_p\equiv x$ mod $p$. 
The definition of the one-loop factor on a non-compact space requires a choice, and we choose here to work with the following expressions:
\begin{align}
&Z^{\mathbb{C}^2}_{\rm 1\mathchar`-loop, \,  bifund }(\vec{a}; \vec{\tilde{a}};m ;\boldsymbol \epsilon)=\prod^N_{\alpha,\beta=1} 
\exp \Big[ \gamma_{\epsilon_1,\epsilon_2}(a_{\alpha}- { \tilde{a}}_{\beta}-m) \Big] \, ,\nonumber \\
&Z^{ A_{p-1}{\rm \mathchar`-ALE}}_{\rm 1\mathchar`-  loop, \, bifund}(\vec{a},\vec{I};\vec{\tilde{a}},\vec{\tilde{I}};m;
\boldsymbol \epsilon):=\prod^N_{\alpha,\beta=1} 
\exp \Big[ \gamma^{(p)}_{\epsilon_1,\epsilon_2}(a_{\alpha}-\tilde{a}_{\beta}-m, \frac{I_{\alpha}-\tilde{I}_{\beta}}{p} ) \Big] \, . \nonumber
\end{align}
The expression for $\mathbb C^2$ agrees with \cite{2003math.....11058N,AGT}, and the one for the ALE space is obtained by the orbifold method.
Then, the $\ell$-factor can be defined as
\begin{equation}
\begin{aligned}
\ell^{
A_{p-1}{\rm \mathchar`-ALE}}_{\rm bifund} (
\vec{a}, 
\boldsymbol{\vec k}
;
\vec{\tilde{a}},
\boldsymbol{\vec{\tilde k}}
;m;\boldsymbol \epsilon
)
&=\frac{ 
\prod^{p-1}_{r=0} 
Z^{\mathbb{C}^2}_{1{\rm \mathchar`-loop, \, bifund}}  (   
\vec{a}^{(r)};\vec{\tilde{a}}^{(r)};m;\epsilon^{(r)}_1 , \epsilon^{(r)}_2
)
 }{
 Z^{A_{p-1}{\rm \mathchar`-ALE}}_{1{\rm \mathchar`-loop , \, bifund} }  
  (   \vec{a} , -{ p}\vec{k}^{(1)};
\vec{\tilde{a}},   -{ p}\vec{\tilde{k}}^{(1)}; m; \boldsymbol \epsilon) } \, , \\
 \ell^{
A_{p-1}{\rm \mathchar`-ALE}}_{\rm vector} (\vec{a},
\boldsymbol{\vec k}
;\boldsymbol \epsilon)
&=1/
\ell^{
A_{p-1}{\rm \mathchar`-ALE}}_{\rm bifund} (\vec{a},
\boldsymbol{\vec k}
;\vec{a},
\boldsymbol{\vec k}
;0;\boldsymbol \epsilon)\,,\\
\prod_{i=1}^{N}
 \ell^{
A_{p-1}{\rm \mathchar`-ALE}}_{\rm anti \mathchar`-fund} (\vec{a}, 
\boldsymbol{\vec k}
;\mu_i;\boldsymbol \epsilon)
&=
\ell^{
A_{p-1}{\rm \mathchar`-ALE}}_{\rm bifund} (
-\vec{\mu}_{\rm anti}, 
\boldsymbol{\vec{0}}
;\vec{a},
\boldsymbol{\vec k}
;0;\boldsymbol \epsilon)\, , \\ 
\prod_{i=1}^{N}
 \ell^{
A_{p-1}{\rm \mathchar`-ALE}}_{\rm fund} (\vec{a},
\boldsymbol{\vec k}
;\mu_{i+N};\boldsymbol \epsilon)
&
=
\prod_{i=1}^{N}
\ell^{
A_{p-1}{\rm \mathchar`-ALE}}_{\rm anti\mathchar`-fund}(
\vec{a},
\boldsymbol{\vec k}
;\epsilon_+ -\mu_{i+N};\boldsymbol \epsilon)\,,
\end{aligned}
\nonumber
\end{equation}
where $\vec{\mu}_{\rm anti}=(\mu_1,\cdots,\mu_N)$ and $k^{(0)}_{\alpha}=k^{(p)}_{\alpha}=0$.\footnote{%
Our convention for $\vec k^{(r)}$ agrees with \cite{Bonelli:201208}. In order to compare with \S\ref{sec:A1-resolved}, we set $p=2$ and $\vec k^{(1)}=-\vec k$.
}

We then define the total $\ell$-factor of $U(N)$ theory with $N_{f} = 2N$:
\begin{equation}
\ell^{A_{p-1}{\rm \mathchar`-ALE}}_{N_{\text F}=2N}( \vec{a},\boldsymbol{\vec k};{\boldsymbol \mu};{\boldsymbol \epsilon}) 
=\ell^{ A_{p-1}{\rm \mathchar`-ALE}}_{\rm vector}(\vec{a},\boldsymbol{ \vec{k}};{\boldsymbol \epsilon})
\prod_{i=1}^N \ell^{ A_{p-1}{\rm \mathchar`-ALE}}_{\rm anti \mathchar`- fund}(\vec{a}, \boldsymbol { \vec{k} };\mu_i;{\boldsymbol \epsilon})
 \prod_{j=N+1}^{2N} \ell^{A_{p-1}{\rm \mathchar`-ALE}}_{\rm fund}(\vec{a}, \boldsymbol{ \vec{k}};\mu_j;{\boldsymbol \epsilon}) \, . \nonumber
\end{equation}
Then the instanton partition function for the sector with holonomy $\vec{I}$ and the first Chern class $\boldsymbol{c}$ on the $A_{p-1}$-ALE spaces is
\begin{equation}
\begin{aligned}
&Z^{ A_{p-1}{\rm \mathchar`- resolved}}_{N_{\text F}=2N,\, {\rm inst } ,\, \boldsymbol{c}\, }(\vec{a},\vec{I}; {\boldsymbol \mu} ;q;\boldsymbol \epsilon)
=
\sum_{\boldsymbol{\vec k}\in \mathcal K(\vec I,\boldsymbol c)}
q^{\frac{1}{2} \sum^N_{\alpha=1} \sum^{p-1}_{r,s=1} k^{(r)}_{\alpha} C_{rs} k^{(s)}_{\alpha} }
\\
&\hspace{30mm}
\times
\ell^{ A_{p-1}{\rm \mathchar`- resolved}}_{N_{\text F}=2N}( \vec{a},
\boldsymbol{\vec k}
;{\boldsymbol \mu};\boldsymbol \epsilon)
\label{ApALEinst}
 \times \prod_{r=0}^{p-1} Z^{\mathbb{C}^2}_{N_{\text F}=2N,\, {\rm inst}   } (\vec{a}^{(r)}, {\boldsymbol \mu};q; \epsilon^{(r)}_1 , \epsilon^{(r)}_2)  \, .
\end{aligned}
\end{equation}
We give an explicit calculation for a few orders of $q$ in \S\ref{app:stacky}.

\section{Proposed relations}
\label{sec:proposals}

For the $\mathcal N=2$ $U(N)$ theory with $N_{\rm F}=2N$ on
the $A_1$-ALE space, we propose the following relation between the instanton partition functions given in (\ref{orb}) and (\ref{A1ALEinst}):
\begin{equation}\label{prop}
\begin{aligned}
&Z ^{A_1 
{\rm \mathchar`- resolved}
}_{ N_{\rm F}=2N, \, {\rm inst}, \, c}( \vec{a},\vec{I} ; \boldsymbol \mu;q ; \epsilon_1,\epsilon_2) \\
& \hspace{3em} =
\left\{
\begin{array}{ll}
 Z ^{
 \mathbb{C}^2/\mathbb{Z}_2
 }_{N_{\rm F}=2N, \, {\rm inst},\, c }(   \vec{a},\vec{I};  \boldsymbol \mu;q ; \epsilon_1,\epsilon_2)
& \text{for } c \geq 0 \, ,\\
 (1-(-1)^N q)^{u_N}  
  Z ^{
   \mathbb{C}^2/\mathbb{Z}_2
   }_{N_{\rm F}=2N, \, {\rm inst},\, c}( -\vec{a},\vec{I}; \boldsymbol{\epsilon_+} - \boldsymbol \mu;q ; \epsilon_1,\epsilon_2 )
  & \text{for }c\leq 0 \, .
\end{array}
\right.
\end{aligned}
\end{equation}
Here  $\boldsymbol{\epsilon_+}=(\epsilon_+,\ldots,\epsilon_+)$ has $2N$ repeated entries of $\epsilon_+=\epsilon_1+\epsilon_2$ and
\begin{equation}
  u_N=\frac{ \epsilon_+(2\sum_i a_i+\sum_{i=1}^{N}\mu_i -
\sum_{j=N+1}^{2N}\mu_j
 )}{2\epsilon_1 \epsilon_2}\,.
\end{equation}
We recall that $\boldsymbol \mu=(\mu_1,\ldots,\mu_N)$ denotes the hypermultiplet masses, $\vec I=(I_1,I_2)$ labels the holonomies at infinity, and $c$ parametrizes the first Chern class.
We checked our proposal (\ref{prop}) for $N\in\{1,\ldots,5\}$, $c\in \{-5,-4,\ldots, +5\}$, and all possible values of holonomies $\vec I$, up to $q^3$.
This is the main result of the paper.

We give examples of the calculations in \S\ref{app:comparison}.
The relation (\ref{prop}) predicts that for $c=0$ the two expressions on the right hand side are equal.
We also observe that the orbifold partition function is invariant under the sign flip of $c$,
$Z^{\mathbb{C}^2/\mathbb{Z}_2}_{N_{\rm F}=2N,{\rm inst}, c }
 = Z^{\mathbb{C}^2/\mathbb{Z}_2}_{N_{\rm F}=2N,{\rm inst}, -c  }$,
for all the terms we calculated although the two sides of the equality involve sums over different sets of Young diagrams.
We expect this property to hold to all orders in~$q$.

We also investigated the $p=3$ case.
When all of $c_{(r)}$ are simultaneously non-negative or non-positive, we found  the following relations for the terms we calculated:
\begin{equation}\label{propAp}
\begin{aligned}
&
Z^{ A_{2} {\rm \mathchar`- resolved}
}_{N_\text F = 2N, \, {\rm inst}, \, \boldsymbol{c} \, }(\vec a, \vec{I} ; \boldsymbol \mu ; q ;  \epsilon_1,\epsilon_2)\\
&
=
\left\{
\begin{array}{ll}
 Z^{\mathbb{C}^2/\mathbb{Z}_3}_{N_\text F = 2N, \, {\rm inst},\, \boldsymbol{c} \, }( \vec a, \vec{I} ; \boldsymbol \mu ;q; \epsilon_1,\epsilon_2)
& \text{for } c_{(r)}\geq 0\ \forall r\,,
\\
 (1- (-1)^N q)^{u^{(2)}_N}  
  Z^{  \mathbb{C}^2/\mathbb{Z}_3 }_{N_\text F = 2N, \, {\rm inst},\, \boldsymbol{c} \, }(
-\vec a,\vec{I}; \, 
   \boldsymbol{\epsilon_+} - \boldsymbol \mu;  
    q;\epsilon_2,\epsilon_1 )  & 
 \text{for } c_{(r)}\leq 0\ \forall r\,,
\end{array}
\right.
\end{aligned}
\end{equation}
where
\begin{equation}
u^{(p-1)}_N=\frac{ \epsilon_+ (2\sum_i a_i+\sum_{i=1}^N \mu_i - \sum_{j=N+1}^{2N}\mu_j)}{p\epsilon_1 \epsilon_2}\,.
\end{equation}
Note the exchange of $\epsilon_1$ and $\epsilon_2$ in the last line of (\ref{propAp});
this is immaterial in the $p=2$ case because the two parameters have the same $\mathbb Z_2$ charge $\pi\sim -\pi$.
For other values of the first Chern class we have not found conclusively such simple relations.
In \S \ref{app:comparison}, we summarize the calculations that we performed to check (\ref{propAp}).
We also checked, for the terms we computed, that the orbifold partition function (\ref{orb}) is invariant under the action of the Weyl group.

\section{Discussion}
\label{sec:discussion}

Our proposals in the previous section immediately raise the question:
are there a pair of distinct two-dimensional theories that naturally correspond to the two schemes?
Obvious candidates for $p=2$ are the two theories discussed in \cite{Belavin:2011sw}, the first one involving two copies of Liouville theory and the second involving super Liouville theory.
The pair of 2d theories naturally generalizes to $p\geq 3$ \cite{Bonelli:201208}.
We leave the study of this question for the future.

The two counting schemes we reviewed in \S \ref{sec:orbifold} and \S\ref{sec:resolved} deal differently with the singularities in the instanton moduli space that appear when the exceptional cycles are blown down.
One might try to interpret the relations (\ref{prop}, \ref{propAp}) as an analog of wall-crossing formulas for equivariant Donaldson invariants (see for example \cite{2006math......6180G} and the references therein).%
\footnote{%
We are very grateful to Y. Tachikawa for suggesting the possible relevance of wall-crossing.
}

Finally, we note that different instanton counting schemes for ALE spaces can be used to define different 't Hooft line operators in four-dimensional ${\mathcal N}=2$ gauge theories.
The correspondence \cite{Kronheimer:MTh} between instantons on a multi-centered Taub-NUT($\sim$ALE) space and monopoles with Dirac singularities can be used to compute the expectation value of a 't Hooft operator on various geometries \cite{Gomis:2011pf,Ito:2011ea,Gang:2012yr}.
Our findings can be adapted for the calculation of 't Hooft operators, which precisely match the predictions from Liouville theory \cite{irreducible}.

\section*{Acknowledgements}

We thank Giulio Bonelli, Yuji Tachikawa, Masato Taki, Alessandro Tanzini, and Futoshi Yagi for valuable discussions
and comments.
The research of Y.I. is supported in part by a JSPS Research Fellowship for Young Scientists.
The research of K.M. is supported in part by a JSPS Postdoctoral Fellowship for Research Abroad.
T.O. is supported in part by the Grant-in-Aid for Young Scientists (B) No. 23740168 and by the Grant-in-Aid for Scientific Research (B) No. 20340048.

\appendix

\section{Explicit calculations}
\subsection{Instantons on $\mathbb{C}^2$}
\label{app:C2}
We review briefly the instanton partition function \cite{Nekrasov:2002qd} for $U(N)$ gauge theory with $N_{\rm F}=2N$ on $\mathbb{C}^2$.
We denote an $N$-tuple of Young diagrams by $\vec{Y}=(Y_1, \cdots, Y_N)\, $. The
instanton number is given by the total number of boxes in the $N$-tuple of Young diagrams $|\vec{Y} | = |Y_1| +\cdots + |Y_N|\,$.  
The contribution of the fixed point labeled by $\vec{Y}$ takes the  form
\begin{align}\label{eq:ZY-C2}
Z^{\mathbb{C}^2}_{\vec{Y}}(
\vec{a}; {\boldsymbol \mu} ; \boldsymbol \epsilon  )=
  \frac{ 
\prod_{i=1}^{N} z_{\rm anti\mathchar`-fund}(\vec{a}; \mu_i ;\boldsymbol\epsilon;\vec{Y} ) 
\prod_{j=N+1}^{2N} z_{{\rm fund}}(\vec{a};\mu_j; \boldsymbol\epsilon;\vec{Y}) 
}{
z_{\rm vector} (\vec{a};\boldsymbol\epsilon;\vec{Y} )
} \, .
 \end{align}
This includes contributions from the vector multiplet as well as the $N$ anti-fundamental and fundamental hypermultiplets; 
they are denoted by $z_{\rm vector}$, $z_{\rm anti\mathchar`-fund}$ and $z_{{\rm fund}}$ respectively. 
These are obtained by taking  products of weights of the equivariant action $(\mathbb{C}^*)^N \times (\mathbb{C}^*)^{2N} \times (\mathbb{C}^*)^2$, whose parameters are $(\vec{a};\boldsymbol \mu;\boldsymbol\epsilon)$. 
 
 Let $Y_{\alpha}=\{\lambda_{\alpha,1}, \lambda_{\alpha,2},\cdots \} \ (1\leq \alpha \leq N)$ be a Young diagram where $\lambda_{\alpha,i}$ is the height of the $i$-the column. We set $\lambda_{\alpha,i}=0$ when $i$ is larger than the width of the diagram $Y_{\alpha}$. Let $Y_{\alpha}^{T}=\{ \lambda^{\prime}_{\alpha,1}, \lambda^{\prime}_{\alpha,2}, \cdots \}$ be its transpose. For a box $s$ in the $i$-the column and the $j$-th row, we define its arm-length $A_{Y_{\alpha}}(s)$ and leg-length $L_{Y_{\alpha}}(s)$ with respect to the diagram $Y_{\alpha}$ by
$A_{Y_{\alpha}}(s)=\lambda_{\alpha,i} - j$, $L_{Y_{\alpha}}(s)=\lambda^{\prime}_{\alpha,j} - i $.
We then define
 \begin{align}
E(a, Y_{\alpha}, Y_{\beta}, s) =a-\epsilon_1 L_{Y_{\beta}}(s) +\epsilon_2 ( A_{Y_{\alpha}}(s) +1) \, .
\end{align}
We set $\epsilon_+=\epsilon_1+\epsilon_2$. 
The contribution from the vector multiplet is \cite{Flume:2002az}
 \begin{align}\label{z-vec}
z_{\rm vector}(\vec{a};\boldsymbol\epsilon; \vec{Y})=\prod_{\alpha,\beta=1}^{N} \displaystyle\prod_{s\in Y_\alpha}  E(a_{\alpha}-a_{\beta}, Y_{\alpha}, Y_{\beta}, s)  \displaystyle\prod_{t\in Y_{\beta}} (\epsilon_+ - E(a_{\beta}-a_{\alpha}, Y_{\beta}, Y_{\alpha}, t) )\, .
 \end{align}
Note that $L_{Y_{\beta}}(s)$ in $E(a_{\alpha}-a_{\beta}, Y_{\alpha}, Y_{\beta}, s)$ is negative when the box $s$ is inside the diagram $Y_{\alpha}$ but outside the diagram $Y_{\beta}\, $. 
The contributions from the fundamental and the anti-fundamental hypermultiplets are given by
\begin{align}
&z_{{\rm fund}}(\vec{a};\mu;\boldsymbol\epsilon;\vec{Y})=\displaystyle\prod_{\alpha=1}^{N} \displaystyle\prod_{s\in Y_{\alpha}} (\phi(a_{\alpha},s) -\mu+\epsilon_1+\epsilon_2) \, , 
\label{z-fund}
\\
&z_{\rm anti\mathchar`-fund}(\vec{a};\mu;\boldsymbol\epsilon;\vec{Y})=
z_{{\rm fund}}(\vec{a};\epsilon_+-\mu;\boldsymbol\epsilon;\vec{Y}) \, , 
\label{z-anti-fund}
\end{align}
where $\phi(a,s)=a+\epsilon_1(i-1)+\epsilon_2(j-1) $
for the box $s$ at the position $(i,j)$ \cite{Bruzzo:2002xf}.

Then the instanton partition function is denoted by
  \begin{align}
Z^{\mathbb{C}^2}_{N_{\rm F}=2N, \, {\rm inst}}(\vec{a}; {\boldsymbol \mu};q;\boldsymbol\epsilon)
= \sum_{n=0}^\infty q^n\, Z^{\mathbb{C}^2}_{ n}(
\vec{a}; {\boldsymbol \mu} ;\boldsymbol\epsilon ) \,,
\end{align}
where $q$ is the one-instanton factor and the coefficient $Z^{\mathbb{C}^2}_{N_{\rm F}=2N, \, n}$
is the sum of contributions $Z^{\mathbb{C}^2}_{\vec{Y}}$ \eqref{eq:ZY-C2} with $|\vec{Y}|=n$.
For example, the coefficient for $N=2$ and $n=1$ is calculated as
\begin{align}
&\quad\
Z^{\mathbb{C}^2}_{ 1} (  \vec{a};{\boldsymbol \mu} ; \boldsymbol\epsilon )  =Z^{\mathbb{C}^2}_{((1),(\varnothing)) } (  \vec{a};{\boldsymbol \mu} ; \boldsymbol\epsilon )  +Z^{\mathbb{C}^2}_{ ((\varnothing),(1))} (  \vec{a};{\boldsymbol \mu} ; \boldsymbol\epsilon ) \nonumber\\
 &
=
\frac{(a_1+\mu_1)(a_1+\mu_2)(a_1+\epsilon_1+\epsilon_2-\mu_3)(a_1+\epsilon_1+\epsilon_2-\mu_4) }{ (-a_1+a_2)\epsilon_1 \epsilon_2 (a_1-a_2+\epsilon_1+\epsilon_2) } 
+ (a_1\leftrightarrow a_2)\,.
\label{zc21}
 \end{align}

\subsection{Orbifolded instantons}
\label{app:orbifold}

We compute explicitly several terms in the orbifold partition function (\ref{orb}).
We expand the partition function as
\begin{align}
Z^{\mathbb{C}^2 / \mathbb{Z}_p }_{N_{\rm F}=2N, \, \boldsymbol{c}\, }(\vec{a},\vec{I}; {\boldsymbol \mu};q;\boldsymbol\epsilon )
=: \sum_n q^n\, Z^{\mathbb{C}^2 / \mathbb{Z}_p}_{\boldsymbol{c}, n}(
\vec{a}, \vec{I};{\boldsymbol \mu} ; \boldsymbol\epsilon) \, . \nonumber
\end{align}
For $p=2$, $N=2$, $\vec{I}=(0,0)$, and $c=-2$, the coefficients with $n=1,2$ are given as
\begin{align}
&\quad\
 Z^{\mathbb{C}^2 / \mathbb{Z}_2}_{c=-2, \, n=1}(
\vec{a},(0,0); {\boldsymbol \mu} ;\boldsymbol \epsilon  )
=
Z^{\mathbb{C}^2 / \mathbb{Z}_2}_{\vec{Y}=((2,1),(\varnothing))}(
\vec{a},(0,0);{\boldsymbol \mu} ;\boldsymbol \epsilon  ) 
+
Z^{\mathbb{C}^2 / \mathbb{Z}_2}_{ \vec{Y}=((\varnothing),(2,1))}(
\vec{a},(0,0);{\boldsymbol \mu} ;\boldsymbol \epsilon  )  \nonumber\\
&
= \frac{(a_1+\mu_1) (a_1+\mu_2) (a_1+\epsilon_1+\epsilon_2-\mu_3) (a_1+\epsilon_1+\epsilon_2-\mu_4)}{(-a_1+a_2) (a_1-a_2+\epsilon_1+\epsilon_2)}
 + (a_1\leftrightarrow a_2)
\label{c1-2n1orb}
\,,
\end{align}
and
\begin{align}
 Z^{\mathbb{C}^2 / \mathbb{Z}_2}_{ c=-2, \, n=2}(
 \vec{a},(0,0); {\boldsymbol \mu} ;\boldsymbol\epsilon   ) 
 &=
 Z^{\mathbb{C}^2 / \mathbb{Z}_2}_{ ((4,1),(\varnothing))}
 +Z^{\mathbb{C}^2 / \mathbb{Z}_2}_{((2,1,1,1),(\varnothing))}
+Z^{\mathbb{C}^2 / \mathbb{Z}_2}_{((2,1),(2))}
 +Z^{\mathbb{C}^2 / \mathbb{Z}_2}_{((2,1),(1,1))}
 +Z^{\mathbb{C}^2 / \mathbb{Z}_2}_{((2),(2,1))}
\nonumber \\
&\quad
+Z^{\mathbb{C}^2 / \mathbb{Z}_2}_{ ((1,1),(2,1))}
+Z^{\mathbb{C}^2 / \mathbb{Z}_2}_{ ((\varnothing),(4,1))}
+Z^{\mathbb{C}^2 / \mathbb{Z}_2}_{ ((\varnothing),(2,1,1,1))}
 \label{c1-2n2orb}
\end{align}
(with the same arguments), where
\begin{align}
& Z^{\mathbb{C}^2 / \mathbb{Z}_2}_{((4,1),(\varnothing))}
=
(a_1+\mu_1) (a_1+2 \epsilon_2+\mu_1) (a_1+\mu_2) (a_1+2 \epsilon_2+\mu_2)  
\times
\frac{(a_1+\epsilon_1+\epsilon_2-\mu_3)}{(-a_1+a_2)}
\nonumber\\
&\hspace{27mm}
\times
\frac{ (a_1+\epsilon_1+3 \epsilon_2-\mu_3) (a_1+\epsilon_1+\epsilon_2-\mu_4) (a_1+\epsilon_1+3 \epsilon_2-\mu_4)}{  (-a_1+a_2-2 \epsilon_2) (\epsilon_1-\epsilon_2) (2\epsilon_2) (a_1-a_2+\epsilon_1+\epsilon_2) (a_1-a_2+\epsilon_1+3 \epsilon_2)} \, ,
\nonumber\\
&
\hspace{-1mm}
Z^{\mathbb{C}^2 / \mathbb{Z}_2}_{ ((2,1,1,1),(\varnothing))}
\hspace{-1mm}
=
(\epsilon_1\leftrightarrow \epsilon_2) \,, \
Z^{\mathbb{C}^2 / \mathbb{Z}_2}_{ ((\varnothing),(4,1))}
\hspace{-1mm}
= (a_1\leftrightarrow a_2)
 \, , \
Z^{\mathbb{C}^2 / \mathbb{Z}_2}_{ ((\varnothing),(2,1,1,1))}
\hspace{-1mm}
=(a_1\leftrightarrow a_2, \,  \epsilon_1\leftrightarrow \epsilon_2)
 \, , \nonumber
\end{align}
and
\begin{align}
&Z^{\mathbb{C}^2 / \mathbb{Z}_2}_{ ((2,1),(2))}
=
(a_1+\mu_1) (a_2+\mu_1) (a_1+\mu_2) (a_2+\mu_2) 
\times
\frac{(a_1+\epsilon_1+\epsilon_2-\mu_3)}{ (-a_1+a_2)(\epsilon_1-\epsilon_2)}
\nonumber\\
&\hspace{30mm}
\times  \frac{ (a_2+\epsilon_1+\epsilon_2-\mu_3) (a_1+\epsilon_1+\epsilon_2-\mu_4) (a_2+\epsilon_1+\epsilon_2-\mu_4)}{  (-a_1+a_2+\epsilon_1-\epsilon_2)(2 \epsilon_2) (a_1-a_2+\epsilon_1+\epsilon_2) (a_1-a_2+2 \epsilon_2)} \, , \nonumber\\
&Z^{\mathbb{C}^2 / \mathbb{Z}_2}_{ ((2,1),(1,1))}
=
(\epsilon_1\leftrightarrow \epsilon_2) \,,\
Z^{\mathbb{C}^2 / \mathbb{Z}_2}_{((2),(2,1))}
=
(a_1\leftrightarrow a_2)\,,\
Z^{\mathbb{C}^2 / \mathbb{Z}_2}_{ ((1,1),(2,1))}
=
(a_1\leftrightarrow a_2, \,  \epsilon_1\leftrightarrow \epsilon_2) \, .  \nonumber
\end{align}
In the case where $p=3$, $N=2$, $\boldsymbol{c}=(-1,-1)$ and $\vec{I}=(2,1)$, the coefficients for $n=\frac{2}{3},\frac{5}{3}$ are given as
\begin{align}
\label{c1-1-1n2/3orb}
&\qquad\qquad
 Z^{\mathbb{C}^2/\mathbb{Z}_3}_{\boldsymbol{c}=(-1,-1) , \, n=\frac{2}{3} }(\vec I=(2,1))= Z^{\mathbb{C}^2/\mathbb{Z}_3}_{ ( (2), (1,1) )}(2,1)=1 \, ,  
\\
Z^{\mathbb{C}^2/\mathbb{Z}_3}_{(-1,-1) , \frac{5}{3} }(2,1)
 &=
 Z^{\mathbb{C}^2/\mathbb{Z}_3}_{  ((2,1,1,1), (1,1)) }(2,1)
  +Z^{\mathbb{C}^2/\mathbb{Z}_3}_{ ( (2,2,1), (1,1) )}
  +Z^{\mathbb{C}^2/\mathbb{Z}_3}_{ ( (2,2,1,1), (1) )}
 +Z^{\mathbb{C}^2/\mathbb{Z}_3}_{  ((2), (1,1,1,1,1)) }  
  \nonumber
\\ &
\qquad
 +Z^{\mathbb{C}^2/\mathbb{Z}_3}_{  ((2), (4,1)) }
 +Z^{\mathbb{C}^2/\mathbb{Z}_3}_{  ((2), (3,2) )} 
  +Z^{\mathbb{C}^2/\mathbb{Z}_3}_{  ((1), (4,2)) }
   +Z^{\mathbb{C}^2/\mathbb{Z}_3}_{ ( (5), (1,1)) }
 \nonumber
\\
  &=\frac{(a_1 + \epsilon_1 + \mu_1) (a_1 + \epsilon_1 + \mu_2) (a_1 + 2 \epsilon_1 + \epsilon_2 - \mu_3) (a_1 + 2 \epsilon_1 + \epsilon_2 -
     \mu_4)}{3 \epsilon_1 (-a_1 + a_2 + \epsilon_1) (a_1 - a_2 + \epsilon_2) (-2 \epsilon_1 + \epsilon_2)} \nonumber\\
     &
\quad
+\frac{ (a_1 + 
    \epsilon_1 + \mu_1) (a_1 + \epsilon_1 + \mu_2) (a_1 + 2 \epsilon_1 + \epsilon_2 - \mu_3) (a_1 + 2 \epsilon_1 + \epsilon_2 - 
    \mu_4)}{(a_1 - a_2 + 2 \epsilon_1) (2 \epsilon_1 - \epsilon_2) (-a_1 + a_2 - \epsilon_1 + \epsilon_2) (-\epsilon_1 + 
    2 \epsilon_2)}
\nonumber
\\
     &\quad 
     +\frac{ (a_1 + \epsilon_1 + \mu_1) (a_1 + \epsilon_1 + \mu_2) (a_1 + 2 \epsilon_1 + \epsilon_2 - 
    \mu_3) (a_1 + 2 \epsilon_1 + \epsilon_2 - \mu_4)}{(-a_1 + a_2 - 2 \epsilon_1) (-a_1 + a_2 - 
    \epsilon_2) (a_1 - a_2 + 3 \epsilon_1 + \epsilon_2) (a_1 - a_2 + \epsilon_1 + 2 \epsilon_2)}  \nonumber\\
     &\quad+
    \frac{ (a_2 + 2 \epsilon_1 + 
    \mu_1) (a_2 + 2 \epsilon_1 + \mu_2) (a_2 + 3 \epsilon_1 + \epsilon_2 - \mu_3) (a_2 + 3 \epsilon_1 + \epsilon_2 - 
    \mu_4)}{3 (a_1 - a_2 - \epsilon_1) \epsilon_1 (-2 \epsilon_1 + \epsilon_2) (-a_1 + a_2 + 2 \epsilon_1 + 
    \epsilon_2)}
 \nonumber
\\
&\quad+ (a_1\leftrightarrow a_2, \epsilon_1\leftrightarrow \epsilon_2)\,.
\label{c1-1-1n5/3orb}
    \end{align}

\subsection{Instantons on the resolved spaces}
\label{app:stacky}
Let us give explicit expressions for several terms in (\ref{A1ALEinst}) and (\ref{ApALEinst}). 
Again we introduce 
\begin{align}
Z^{A_{p-1}{\rm \mathchar`- resolved} }_{N_{\rm F}=2N, \, \boldsymbol{c} \, }(\vec{a},\vec{I}; {\boldsymbol \mu},q ; \boldsymbol \epsilon )
=: \sum_n q^n\, Z^{A_{p-1}{\rm \mathchar`- resolved} }_{ \boldsymbol{c}, \, n}(
\vec{a},\vec{I}; {\boldsymbol \mu} ; \boldsymbol \epsilon ) \, .
\end{align}
for the coefficients of $q^n$.
For  $p=2,\, N=2,\, \vec{I}=(0,0)$ and $c_1=-2$, the coefficients for $n=1,2$ computed from (\ref{A1ALEinst}) are
\begin{align}
&\quad Z^{A_1{\rm \mathchar`- resolved} }_{ c=-2, \, n=1}(
\vec{a} , (0,0);  {\boldsymbol \mu}, {\boldsymbol \epsilon} )
=\ell^{A_1{\rm \mathchar`- resolved} }_{N_{\rm F}=4} \Big( \vec{k}=(1,0) \Big)
+\ell^{A_1{\rm \mathchar`- resolved} }_{N_{\rm F}=4} \Big( \vec{k}=(0,1) \Big) \nonumber\\
& =\frac{
(-a_1+\epsilon_1+\epsilon_2-\mu_1)
(-a_1+\epsilon_1+\epsilon_2-\mu_2)
(-a_1+\mu_3)
(-a_1+\mu_4)
}
{
(a_1-a_2)(-a_1+a_2+\epsilon_1+\epsilon_2)
} 
+(a_1\leftrightarrow a_2)
 \label{c1-2n1stacky}
\, ,  
\\
&\quad\
Z^{A_1{\rm \mathchar`- resolved} }_{ c=-2, \, n=2}(
\vec{a}, (0,0);{\boldsymbol \mu} ; \boldsymbol\epsilon )
\nonumber
\\
&
 =\ell^{A_1}  (1,0)
\Big(  Z^{\mathbb{C}^2}_{ n=1}(
a_1- 2 \epsilon_1,a_2;
 {\boldsymbol \mu};
2\epsilon_1,\epsilon_2-\epsilon_1
  )
 +  Z^{\mathbb{C}^2}_{ n=1}(
a_1- 2 \epsilon_2,a_2;
 {\boldsymbol \mu};
\epsilon_1-\epsilon_2, 2\epsilon_2  )\Big) 
\nonumber\\
&\qquad
 +\ell^{A_1 } (0,1) 
\Big(  Z^{\mathbb{C}^2}_{ n=1}(
a_1,a_2- 2 \epsilon_1;
 {\boldsymbol \mu};
2\epsilon_1,\epsilon_2-\epsilon_1
  )
 + 
Z^{\mathbb{C}^2}_{ n=1}(
a_1, a_2-2 \epsilon_2;{\boldsymbol \mu};
\epsilon_1-\epsilon_2, 2\epsilon_2  ) \Big) \nonumber
\\
&=
\frac{(-a_1+\epsilon_1+\epsilon_2-\mu_1) (-a_1+\epsilon_1+\epsilon_2-\mu_2) (-a_1+\mu_3) (-a_1+\mu_4)}{(a_1-a_2) (-a_1+a_2+\epsilon_1+\epsilon_2)}
 \nonumber
\\ &\hspace{2em}
 \times
\Big(
\frac{(a_1-2 \epsilon_1+\mu_1) (a_1-2 \epsilon_1+\mu_2) (a_1-\epsilon_1+\epsilon_2-\mu_3) (a_1-\epsilon_1+\epsilon_2-\mu_4)}{2 \epsilon_1 (-a_1+a_2+2 \epsilon_1) (-\epsilon_1+\epsilon_2) (a_1-a_2-\epsilon_1+\epsilon_2)}
  \nonumber\\
 &\hspace{4em}
+
\frac{(a_2+\mu_1) (a_2+\mu_2) (a_2+\epsilon_1+\epsilon_2-\mu_3) (a_2+\epsilon_1+\epsilon_2-\mu_4)}{2 (a_1-a_2-2 \epsilon_1) \epsilon_1 (-\epsilon_1+\epsilon_2) (-a_1+a_2+3 \epsilon_1+\epsilon_2)}
\nonumber
\\
 &\hspace{4em} \label{c1-2n2stacky}
+
\frac{(a_1-2 \epsilon_2+\mu_1) (a_1-2 \epsilon_2+\mu_2) (a_1+\epsilon_1-\epsilon_2-\mu_3) (a_1+\epsilon_1-\epsilon_2-\mu_4)}{2 (\epsilon_1-\epsilon_2) (a_1-a_2+\epsilon_1-\epsilon_2) \epsilon_2 (-a_1+a_2+2 \epsilon_2)}
\\
 &\hspace{4em}
+\frac{(a_2+\mu_1) (a_2+\mu_2) (a_2+\epsilon_1+\epsilon_2-\mu_3) (a_2+\epsilon_1+\epsilon_2-\mu_4)}{2 (a_1-a_2-2 \epsilon_2) (\epsilon_1-\epsilon_2) \epsilon_2 (-a_1+a_2+\epsilon_1+3 \epsilon_2)}
\Big)
+ (a_1\leftrightarrow a_2)\,.\nonumber
 \end{align}

For $p=3$, $N=2$, $\vec{I}=(2,1)$ and $\boldsymbol{c}=(-1,-1)$, we compute the coefficients for $n=\frac{2}{3},\frac{5}{3}$ using (\ref{ApALEinst}). 
The coefficient for $n=\frac{2}{3}$ is
 \begin{align}
\label{c1-1-1n2/3stacky}
 &Z^{  A_2{\rm \mathchar`-resolved}}_{ \boldsymbol{c}=(-1,-1) , \, n=\frac{2}{3} }(2,1) = 
 \ell_{N_{\rm F}=4}^{A_2}  \left( \vec{a},\boldsymbol{\vec k}
 =
  \{
 (-\frac{2}{3},-\frac{1}{3}),(-\frac{1}{3},-\frac{2}{3})
\}  ; {\boldsymbol \mu} \right)\, .
\end{align}
First we focus on the $\ell$ factor for the vector multiplet.
\begin{align}
\ell_{\rm vector} ( \vec{a}, 
\boldsymbol{\vec k}
 ;{\boldsymbol \mu}
 )
 =1\Big/
 \prod_{\alpha,\beta=1}^2
\exp\Bigg[  \Big(
\sum^2_{r=0} \gamma_{\epsilon^{(r)}_1,\epsilon^{(r)}_2}(a^{(r)}_{\alpha \beta} )
\Big)
 - \gamma^{ (3)}_{\epsilon_1,\epsilon_2}(a_{\alpha \beta}\, , - k^{(1)}_{\alpha\beta}) \, 
 \Bigg] \, . \nonumber
\end{align}
For $\boldsymbol{\vec k}
 \hspace{-0.3em}
 =
  \hspace{-0.3em}
  \{
 (-\frac{2}{3},-\frac{1}{3}),(-\frac{1}{3},-\frac{2}{3})
\}$, the exponent for $(\alpha,\beta)=(1,2)$ in the denominator above is 
 \begin{align}
&\quad
\Big(
\sum^2_{r=0} \gamma_{\epsilon^{(r)}_1,\epsilon^{(r)}_2}(a^{(r)}_{12} )
\Big)
 - \gamma^{ (3)}_{\epsilon_1,\epsilon_2}(a_{12}\, , - k^{(1)}_{12})
 \nonumber
\\
 &=
 \frac{d}{ds}\Big|_{s=0} \frac{
         1}{\Gamma(s)}
         \int_0^\infty \frac{dt}{t}
           t^s
               \Bigg[
            e^{-t a_{12}}
                  \Big(
           \frac{X}{(X^3-1)(X^{-2}Y-1)}+  \frac{X^{-1} Y }{ (X^2Y^{-1}-1)(X^{-1}Y^2-1)} \nonumber\\
           \label{ab=12}
           & \hspace{5em}  + \frac{Y^{-1}}{(XY^{-2}-1)(Y^3-1)} 
            - \frac{X+X^2Y+Y^2}{(X^3-1)(Y^3-1)}       \Big)    \Bigg] 
 =0 \, ,
                        \end{align}
where $X:=e^{t\epsilon_1}$ and $Y:=e^{t\epsilon_2}$.                       
By combining this with the factors for $(\alpha,\beta)=(2,1)$ as well as with those for hypermultiplets, we get $ \ell^{A_2}_{N_{\rm F}=4}=1$ for
$\boldsymbol{\vec k}  =  \{ (-\frac{2}{3},-\frac{1}{3}),(-\frac{1}{3},-\frac{2}{3})\}$, so (\ref{c1-1-1n2/3stacky})=$1$.

 The coefficient for $n=\frac{5}{3}$ is
\begin{align}
Z^{  A_2{\rm \mathchar`-resolved}}_{  \boldsymbol{c}=(-1,-1) , \, n=\frac{5}{3} }(2,1)&  =\ell^{A_2}_{N_{\text F}=4} \left( \{ (-\frac{2}{3},-\frac{1}{3}),(-\frac{4}{3},\frac{1}{3})\}\right )
+\ell^{A_2}_{N_{\text F}=4} \left (\{ (\frac{1}{3},-\frac{4}{3}),(-\frac{1}{3},-\frac{2}{3})\} \right )
\nonumber
\\
&\quad
 +\ell^{A_2}_{N_{\text F}=4} \left(\{ (-\frac{2}{3},-\frac{1}{3}),(-\frac{1}{3},-\frac{2}{3})\} \right)  
  \sum^2_{r=0} Z^{\mathbb{C}^2}_{ 1} (  \vec{a}^{(r)};{\boldsymbol \mu}; \epsilon^{(r)}_1,\epsilon^{(r)}_2 )  \, ,
  \label{c1-1-1n5/3stacky}
 \end{align}
 where $\vec{a}^{(r)},\epsilon_{1,2}^{(r)}$ are the ones for $\boldsymbol{\vec k}
 \hspace{-0.3em}
 =
  \hspace{-0.3em}
  \{
 (-\frac{2}{3},-\frac{1}{3}),(-\frac{1}{3},-\frac{2}{3})
\} 
$ and $Z^{\mathbb{C}^2}_{N_{\rm F}=4,  \, 1}(  \vec{a};{\boldsymbol \mu} ; \boldsymbol \epsilon )$ is given in (\ref{zc21}).
 As in (\ref{ab=12}), let us focus on one exponential in the $\ell$ factor for the vector multiplet. For $\boldsymbol{\vec k}
 \hspace{-0.3em}
 =
  \hspace{-0.3em}
  \{
 (-\frac{2}{3},-\frac{1}{3}),(-\frac{4}{3},\frac{1}{3})
\}$, it has the exponent
  \begin{align}
&\Big(
\sum^2_{r=0} \gamma_{\epsilon^{(r)}_1,\epsilon^{(r)}_2}(a^{(r)}_{12} )
\Big)
 - \gamma^{ (3)}_{\epsilon_1,\epsilon_2}(a_{12}, -k^{(1)}_{12})
 \nonumber\\
 &=
 \frac{d}{ds}\Big|_{s=0} \frac{
         1}{\Gamma(s)}
         \int_0^\infty \frac{dt}{t}
           t^s
               \Bigg[
            e^{-t a_{12}}
                  \Big(
           \frac{X}{(X^3-1)(X^{-2}Y-1)}+  \frac{X^3 Y^{-1}}{(X^2Y^{-1}-1)(X^{-1}Y^2-1)} \nonumber\\
           & \qquad + \frac{Y^5}{(XY^{-2}-1)(Y^3-1)} 
            - \frac{X+X^2Y+Y^2}{(X^3-1)(Y^3-1)}
           \Big)  \Bigg] =
\log \Big[ (a_{12}-\epsilon_1 )( a_{12} -2\epsilon_2 )     \Big] \, . \nonumber
 \end{align}
 Combining it with other factors, we obtain
   \begin{equation}
 \ell_{\rm vector}\left (\vec{a},   \{ (-\frac{2}{3},-\frac{1}{3}),(-\frac{4}{3},\frac{1}{3})\} \right)
=\frac{1}{ (a_{12}-\epsilon_1 )( a_{12} -2\epsilon_2 )       ( a_{21} +2\epsilon_1+\epsilon_2 )                                        ( a_{21} +\epsilon_1+3\epsilon_2  )}\,.\nonumber
   \end{equation}
Similarly we obtain
\begin{align}
 &
 \prod_{i=1}^2
 \ell_{\rm anti \mathchar`-fund}\left (\vec{a}, 
   \{
 (-\frac{2}{3},-\frac{1}{3}),(-\frac{4}{3},\frac{1}{3})
\} ;\mu_i \right )
=(a_1+\mu_1-\epsilon_1-2\epsilon_2)(a_1+\mu_2-\epsilon_1-2\epsilon_2) \, , \nonumber
\\
 &
 \prod_{i=3}^4
 \ell_{\rm fund}\left (\vec{a},
   \{
 (-\frac{2}{3},-\frac{1}{3}),(-\frac{4}{3},\frac{1}{3})
\} ;\mu_i\right)
=(a_1-\mu_3-\epsilon_2)(a_1-\mu_4-\epsilon_2) \, , \nonumber
\end{align}
\begin{align}
&
 \ell_{\rm vector}\left(\vec{a}, 
    \{ (\frac{1}{3},-\frac{4}{3}),(-\frac{1}{3},-\frac{2}{3})
\} \right )
=
 \frac{1}{  (a_{12}+3\epsilon_1+\epsilon_2) (a_{12} + \epsilon_1+2\epsilon_2 )( a_{21} -2\epsilon_1)                      ( a_{21}-\epsilon_2  )  } \,,
\nonumber
\\
&
\prod_{i=1}^2 \ell_{\rm anti \mathchar`-fund}\left(\vec{a},     \{  (\frac{1}{3},-\frac{4}{3}),(-\frac{1}{3},-\frac{2}{3}) \}   ;
\mu_i    \right ) 
=(a_2+\mu_1-2\epsilon_1-\epsilon_2)(a_2+\mu_2-2\epsilon_1-\epsilon_2) \, , 
\nonumber
\\
&\prod_{i=3}^4 \ell_{\rm fund}\left(\vec{a},    \{  (\frac{1}{3},-\frac{4}{3}),(-\frac{1}{3},-\frac{2}{3}) \}   ;
\mu_i     \right) =(a_2-\mu_3-\epsilon_1)(a_2-\mu_4-\epsilon_1) \, . \nonumber
 \end{align}
By collecting these $\ell$ factors, we can calculate (\ref{c1-1-1n5/3stacky}).

\subsection{Comparison}
\label{app:comparison}
Let us compare instanton partition functions computed in the two schemes.
For $p=2$, $N=2$, $\vec{I}=(0,0)$, and $c=-2$ the expressions in (\ref{c1-2n1stacky},\ref{c1-2n2stacky}) and (\ref{c1-2n1orb},\ref{c1-2n2orb}) are related as
  \begin{align}
&Z^{A_1{\rm \mathchar`-resolved} }_{ c=-2, \, n=1}(
\vec{a},\vec 0; {\boldsymbol \mu} ;\boldsymbol \epsilon ) =
 Z^{ \mathbb{C}^2/\mathbb{Z}_2 }_{ c=-2, \, n=1}(
-\vec{a},\vec 0;  \boldsymbol{\epsilon_+} -{\boldsymbol  \mu} ; \boldsymbol\epsilon )\,,\nonumber
\end{align}
\begin{align}
Z^{A_1{\rm \mathchar`- resolved} }_{c=-2,n= 2}(
\vec{a},\vec 0; {\boldsymbol \mu} ;\boldsymbol \epsilon ) 
&
=
 Z^{ \mathbb{C}^2/\mathbb{Z}_2 }_{ -2, 2}(
-\vec{a},\vec 0;\boldsymbol{\epsilon_+} -{\boldsymbol  \mu} ; \boldsymbol\epsilon ) 
\nonumber\\
&\quad
- \frac{
\epsilon_+(2a_1+2a_2+\mu_1+\mu_2-\mu_3-\mu_4)
}{2\epsilon_1\epsilon_2}  
Z^{ \mathbb{C}^2/\mathbb{Z}_2 }_{ -2,1}(
-\vec{a},\vec 0;\boldsymbol{\epsilon_+} -{\boldsymbol  \mu} ; \boldsymbol\epsilon ) \, . \nonumber
 \end{align}
 These are consistent with our proposal (\ref{prop}).

 In the case with $p=3$, $N=2$, $\vec{I}=(2,1)$, $\boldsymbol{c}=(-1,-1)$,  the expressions (\ref{c1-1-1n2/3stacky},\ref{c1-1-1n5/3stacky}) and (\ref{c1-1-1n2/3orb}, \ref{c1-1-1n5/3orb}) are related as
 \begin{align}
&Z^{A_2{\rm \mathchar`- resolved} }_{ \boldsymbol{c}=(-1,-1), \, n=\frac{2}{3}}(
\vec{a},(2,1); {\boldsymbol \mu} ; \epsilon_1,\epsilon_2 ) =
 Z^{ \mathbb{C}^2/\mathbb{Z}_3 }_{ \boldsymbol{c}=(-1,-1), \, n=\frac{2}{3}}(
-\vec{a} ,(2,1);\boldsymbol{\epsilon_+} -{\boldsymbol  \mu} ;  \epsilon_2,\epsilon_1 ) \,, \nonumber\\
&Z^{A_2{\rm \mathchar`- resolved} }_{ \boldsymbol{c}=(-1,-1), \, n=\frac{5}{3}}(
\vec{a},(2,1); {\boldsymbol \mu} ;\epsilon_1,\epsilon_2 ) 
=
 Z^{ \mathbb{C}^2/\mathbb{Z}_3 }_{ \boldsymbol{c}=(-1,-1), \, n=\frac{5}{3}}(
-\vec{a},(2,1);\boldsymbol{\epsilon_+} -{\boldsymbol  \mu} ;  \epsilon_2,\epsilon_1   ) \nonumber\\
&\hspace{2em}- \frac{
\epsilon_+(2a_1+2a_2+\mu_1+\mu_2-\mu_3-\mu_4)
}{3\epsilon_1\epsilon_2}  
Z^{ \mathbb{C}^2/\mathbb{Z}_3 }_{ \boldsymbol{c}=(-1,-1), \, n=\frac{2}{3}}(
-\vec{a},(2,1);\boldsymbol{\epsilon_+} -{\boldsymbol  \mu} ;  \epsilon_2,\epsilon_1  ) \, . \nonumber
 \end{align}
Again these relations confirm (\ref{propAp}).

More generally, we performed the following comparison.
\begin{itemize}
\item For the $U(N)$ theory on the $A_1$ space, we confirmed (\ref{prop}) up to $q^4$ for $N=1,2,3$, and up to $q^3$ for $N=4,5$, with all possible holonomies $\vec I$ and $-5\leq c\leq 5$.

\item For the $U(2)$ theory on the $A_2$ space, we confirmed (\ref{propAp}) up to $q^2$ for $(c_{(1)},c_{(2)})\in \{\pm (1,1),\pm (0,1),\pm (1,0)
\}$ with all possible holonomies $\vec I$.

\item For the $U(3)$ theory on the $A_2$ space, we checked (\ref{propAp}) up to $q^2$ for $(\boldsymbol c;\vec I)=(1,1;0,0,0)$, $(1,1;0,1,2), (-1,-1;0,0,0), (-1,-1;0,1,2), (-1,-2;1,1,2),(-2,-1;0,1,1)$.

\end{itemize}

\bibliography{blowup-orbifold_draft}

\end{document}